\documentclass[a4paper]{jpconf}
\usepackage{graphicx}
\begin{document}
\title{The pygmy dipole strength, the neutron radius
of ${}^{208}$Pb and the symmetry energy}

\author{X. Roca-Maza$^{1,2}$, 
        M. Brenna$^{1,3}$, 
        M. Centelles$^2$, 
        G. Col\`o$^{1,3}$, 
        K. Mizuyama$^{1}$, 
        G. Pozzi$^{3}$,  
        X. Vi\~nas$^2$ and 
        M. Warda$^{2,4}$}

\address{$^1$INFN, sezione di Milano, via Celoria 16, I-20133 Milano, Italy}
\address{$^2$Departament d'Estructura i Constituents de la Mat\`eria
         and Institut de Ci\`encies del Cosmos,
         Facultat de F\'{\i}sica, Universitat de Barcelona,
         Diagonal 647, 08028 Barcelona, Spain}
\address{$^3$Dipartimento di Fisica, Universit\`a degli 
         Studi di Milano, via Celoria 16, I-20133 Milano, Italy}
\address{$^4$Katedra Fizyki Teoretycznej, Uniwersytet Marii Curie--Sk\l odowskiej,
         ul. Radziszewskiego 10, 20-031 Lublin, Poland}

\ead{xavier.roca.maza@mi.infn.it}

\begin{abstract}
The accurate characterization of the nuclear symmetry energy and its
density dependence is one of the outstanding open problems in nuclear
physics. A promising nuclear observable in order to constrain the density
dependence of the symmetry energy at saturation is the neutron skin
thickness of medium and heavy nuclei. Recently, a low-energy peak in the
isovector dipole response of neutron-rich nuclei has been discovered that
may be correlated with the neutron skin thickness. The existence of this
correlation is currently under debate due to our limited experimental
knowledge on the microscopic structure of such a peak. We present a
detailed analysis of Skyrme Hartree-Fock (HF) plus random phase approximation 
(RPA) predictions for the dipole response in several neutron-rich nuclei
and try to elucidate whether models of common use in nuclear physics
confirm or dismiss its possible connection with the neutron skin
thickness. Finally, we briefly present theoretical results for parity
violating electron scattering on ${}^{208}$Pb at the conditions of the PREx
experiment and discuss the implications for the neutron skin thickness of
${}^{208}$Pb and the slope of the symmetry energy.
\end{abstract}

\section{Introduction}
\label{introduction}
Experimental studies \cite{ryez2002, adri2005, weil2009} on the low-energy 
isovector dipole response or Pygmy Dipole Resonance (PDR) in neutron rich nuclei 
are of crucial relevance because they determine reaction rates in the  
$r$-process \cite{paar2007} and, in addition, the PDR has been related to the neutron 
skin thickness of the studied nuclei in Refs.~\cite{klim2007,carb2010,inak2011}. However, the existence of 
this correlation is currently under debate \cite{rein2010} due to our limited 
experimental knowledge of the microscopic structure of such a peak. If the 
underlying dynamics giving rise to the PDR is, eventually, confirmed to be 
strongly correlated to the formation of a neutron skin in neutron-rich nuclei, the 
density dependence of the nuclear symmetry energy at saturation will be immediately 
constrained \cite{brow2000}. This is of broad interest since the symmetry energy and 
its density dependence impact on a variety of physical systems, such as the
composition and structure of the crust in a neutron star \cite{horo2001,roca2008a}, 
the neutron skin thickness of a heavy nucleus \cite{brow2000,cent2009,ward2009}, 
atomic-parity violation \cite{silt2005} and heavy ion collisions \cite{tsan2009}.

For all these reasons, we have performed in Ref.~\cite{pozzi2011} a detailed analysis 
of Skyrme Hartree-Fock (HF) plus random phase approximation (RPA) predictions for 
the isovector and isoscalar dipole response in ${}^{68}$Ni, ${}^{132}$Sn and ${}^{208}$Pb 
nuclei ---representative of different mass regions--- in order to elucidate the nature 
and possible connection of the PDR with the slope of the symmetry energy. The strategy 
adopted to understand if such a connection may exist, is very simple. Giant resonances are 
collective excitations of atomic nuclei that have allowed, in the past, to determine 
some nuclear saturation properties such as the nuclear incompressibility (strongly 
related to the Giant Monopole Resonance \cite{colo2004}), 
the effective nucleon mass (which affects the Giant Quadrupole Resonance \cite{rein1999}), 
or the nuclear symmetry energy at sub-saturation density (which act as a restoring force 
in the Giant Dipole Resonance \cite{trip2008}). Hence, collective phenomena inform us about 
general properties of the nuclear effective interaction and all realistic models should predict  
such a collectivity even though some differences on details may appear. The additional fact 
that the energy weighted sum rule for the PDR has been correlated with the $L$ parameter 
\cite{carb2010}, defined as $L = 3\rho_0 \left.\frac{\partial c_{\rm sym}(\rho)}{\partial \rho}\right\vert_{\rho_0}$  
where $c_{\rm sym}(\rho)$ is the symmetry energy, $\rho$ the baryon density and $\rho_0$ the 
nuclear saturation density, has motivated our selection of studied interactions. Specifically,  
we use three Skyrme interactions widely used for nuclear structure calculations and that 
differ in their predictions of the $L$ parameter to study the collectivity 
displayed by the RPA states giving rise to the pygmy dipole strength, or {\it RPA-pygmy} state, 
as well as its possible relation with the neutron skin thickness. Based on our experience, we 
present here the common features found in different mass regions by using ${}^{208}$Pb as a 
representative example. 

Currently, the PREx collaboration \cite{prexIa} aims to determine the neutron
radius of ${}^{208}$Pb within a 1\% error by parity violating electron scattering
(PVES) \cite{prexIb}. Such a measurement is very important for three basic reasons.
First, it measures the neutron distribution in a heavy nucleus free from
most of the strong interaction uncertainties. Second, it paves the way for
further measurements of neutron densities by PVES \cite{prexII,ban_2011}.
Third, it may allow one to derive a significant constraint on the $L$
parameter \cite{roca2011,cent2010} and, therefore, it may help in constraining the
isovector channel of the nuclear effective interaction. Here, we will shortly
discuss mean-field model predictions for PVES at the kinematics of PREx \cite{roca2011}. 

The work is organized as follows. In Sec.~\ref{formalism} we briefly present the basic 
formalism employed in our analysis. For further information we address the reader to 
Refs.~\cite{prexIb,ring1980}. In Sec.~\ref{results}, we show the main results of our works 
\cite{pozzi2011,roca2011}. Finally, our conclusions are laid in Sec.~\ref{conclusions}. 

\section{Formalism}
\label{formalism}
\subsection{Random Phase Approximation}
\label{formalism-rpa}
The RPA method is well-known from textbooks \cite{ring1980}. 
In short, once the HF equations are solved self-consistently for the given 
Hamiltonian, we build accordingly the residual interaction ---considered to 
be a small perturbation of the HF mean-field potential--- and, then, we solve 
the RPA coupled equations by means of the matrix formulation. 
The important quantities to define for our study are the following. The main one 
is the reduced transition strength or probability, 
\begin{equation}
B(EJ, \tilde 0 \rightarrow \nu) \equiv
\left\vert \sum_{ph} A_{ph}(EJ,\tilde 0 \rightarrow \nu)\right\vert^2=
\left\vert \sum_{ph}
\left( X_{ph}^{(\nu)}
+ Y_{ph}^{(\nu)} \right) \langle p \vert\vert \hat F_{J}
\vert\vert h \rangle \right\vert^2
\label{rts}
\end{equation}
where $A_{ph}(EJ,\tilde 0 \rightarrow \nu)$ is the reduced amplitude, $\vert\tilde{0}\rangle$ 
is the RPA ground state, $\vert\nu\rangle$ is a generic RPA excited state, $J$ is the angular 
momentum carried by the operator, $\hat F_{JM}$, meant to modelize the experimental 
probe and $\langle p \vert\vert \hat F_{J} \vert\vert h \rangle$ is the reduced matrix element 
of such an operator between a hole ($h$) state (occupied state) and a particle ($p$) state 
(unoccupied state). The sum of all $ph$ states that contribute to an RPA transition is weighted 
by the $X^{\nu}$ and $Y^{\nu}$ amplitudes, eigenvectors of the RPA secular matrix \cite{ring1980}. 
For further details we also refer to \cite{colo2011}. 

\subsection{Parity Violating Elastic Electron Scattering}
\label{formalism-pves}
Parity violating electron-nucleus scattering (PVES) probes neutrons in a nucleus 
via the electroweak interaction \cite{donn1989,horo1998}. Electrons interact with 
the protons and neutrons of the nucleus by exchanging a photon or a $Z^0$ boson. 
The former mainly couples to protons while the latter basically couples to neutrons. 
In the experiment, one measures the parity-violating asymmetry,     
\begin{equation}
A_{pv}\equiv
\frac{\displaystyle \frac{d\sigma_+}{d\Omega}-\frac{d\sigma_-}{d\Omega}}
{\displaystyle \frac{d\sigma_+}{d\Omega}+\frac{d\sigma_-}{d\Omega}} ,
\label{apv}
\end{equation}
where $d\sigma_\pm/d\Omega$ is the elastic electron-nucleus differential cross section 
for incident electrons with positive and negative helicity states. For
a realistic calculation of the parity violating asymmetry, we solve the Dirac 
equation via the distorted wave Born approximation (DWBA) \cite{roca2008b} where the main 
input are the electric and weak charge distributions of the studied target 
\cite{roca2011,cent2010}.

\section{Results}
\label{results}
\subsection{Low-energy dipole response of ${}^{208}$Pb}
In our recent work \cite{pozzi2011} we have studied in detail the dipole 
response of ${}^{68}$Ni, ${}^{132}$Sn and ${}^{208}$Pb by means of the formalism explained 
in Section \ref{formalism-rpa} and by using different Skyrme interactions. Here we will present 
the case of ${}^{208}$Pb as a representative example of the common trends found in different 
mass regions. 

\begin{figure}[h!]
\begin{center}
\includegraphics[width=0.45\linewidth,angle=0,clip=true]{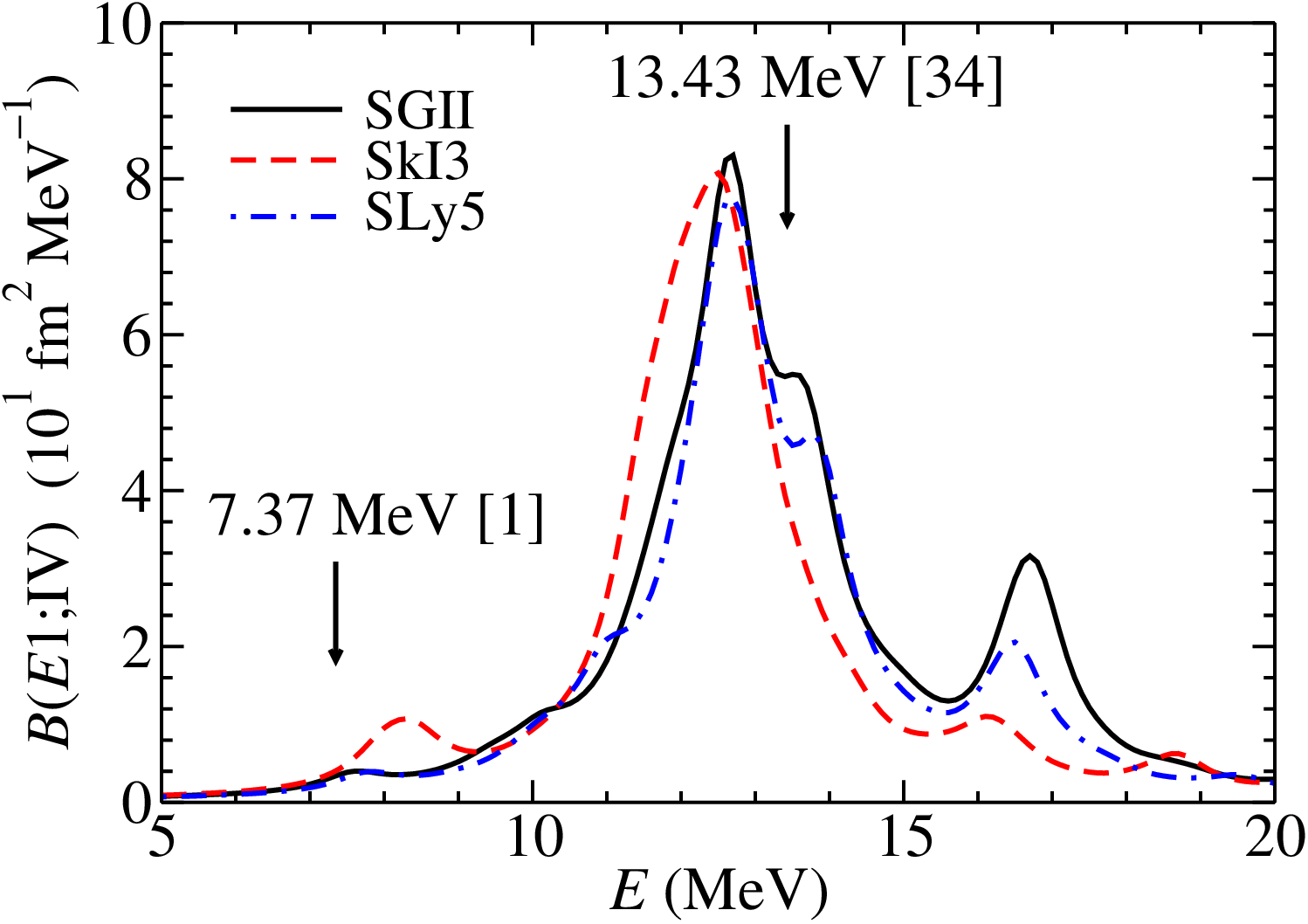}
\includegraphics[width=0.45\linewidth,angle=0,clip=true]{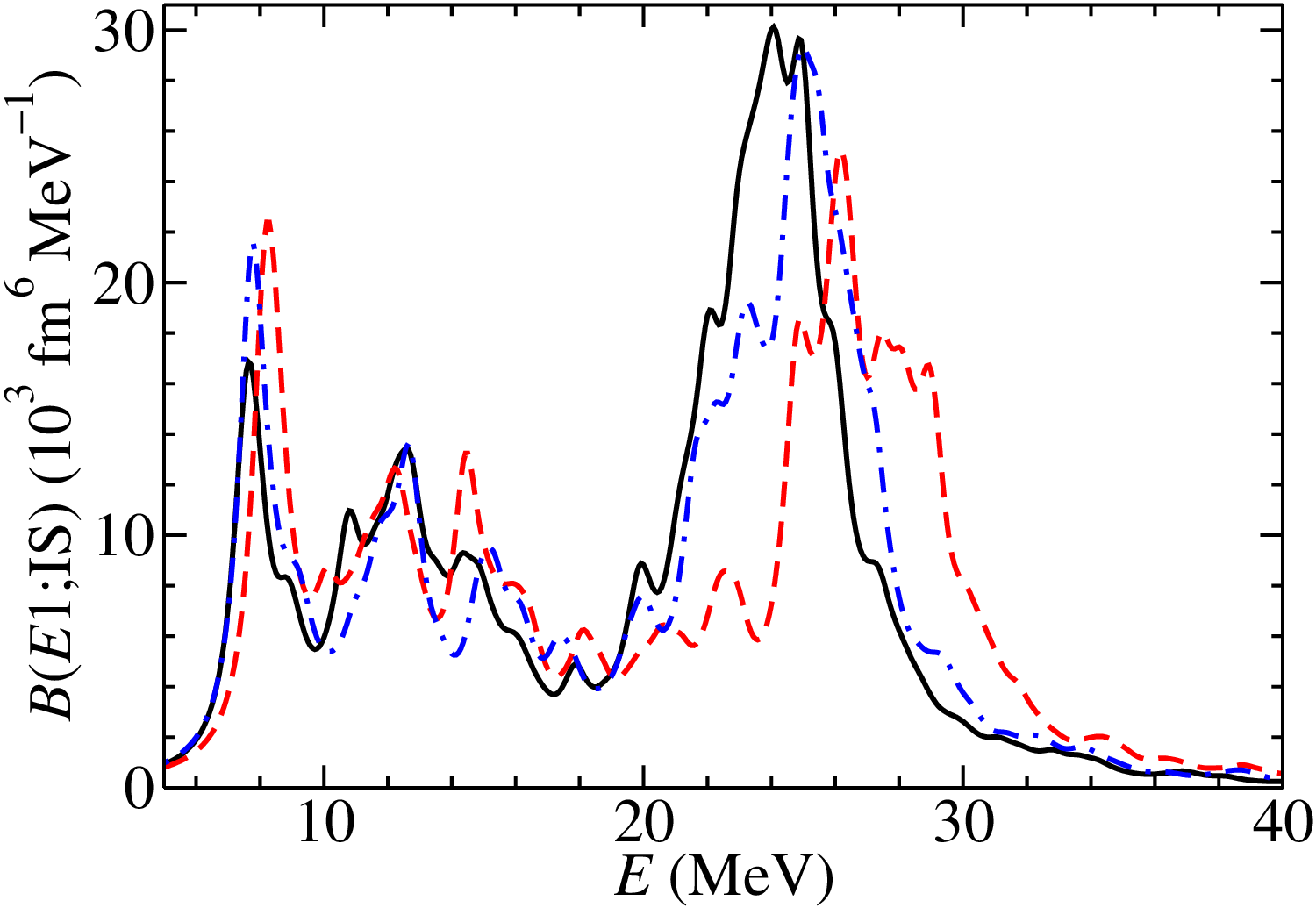}
\end{center}
\caption{\label{fig1} Strength function calculated by convoluting the corresponding reduced transition 
probability [Eq.~(\ref{rts})] with a Lorenzian of 1 MeV width for the isovector (left) and isoscalar 
(right) dipole response of ${}^{208}$Pb as a function of the excitation energy \cite{pozzi2011}. In both figures the 
predictions of SGII, SkI3 and SLy5 are depicted. Black arrows indicate the experimental centroid 
energies for the PDR (E = 7.37 MeV) \cite{ryez2002} and for the IVGDR (E = 13.43 MeV) \cite{berm1975}.}
\end{figure}

In Fig.~\ref{fig1} we show the isovector (IV: left panel) and isoscalar (IS: right panel) 
averaged strength functions for the dipole in ${}^{208}$Pb predicted by the 
Skyrme models SGII \cite{sgii} with an $L=37.6$ MeV, SLy5 \cite{sly5} with an $L=48.3$ MeV and SkI3 
\cite{ski3} with an $L=100.5$ MeV as a function of the excitation energy. Experimental data for the 
centroid energies of the PDR \cite{ryez2002} and the IV Giant Dipole Resonance (GDR) \cite{berm1975} 
are also depicted (black arrows). The theoretical predictions lie within the experimental error and, 
therefore, the Skyrme-HF plus RPA approach may constitute a good starting point for a detailed 
analysis of the microscopic structure of the PDR in those nuclei. An interesting feature that can be also 
observed in different mass regions \cite{pozzi2011} is the ratio between the low- and high-energy  
strength exhausted by the IS peaks: of the same order; and by the IV peaks: 
one order of magnitude. This already indicates the importance of the probe used to 
excite a certain RPA state and reveals also its nature. In other words, the IS dipole operator 
excite more efficiently the {\it RPA-pygmy} state than the IV dipole operator. This means that a perfect
probe for exciting such a state would be mostly isoscalar.   
Moreover, we see from both figures that as the predicted value for the $L$ parameter increases ---in going 
from SGII to SLy5 and, finally, to SkI3, the low-energy peak is shifted to larger excitations energies and 
to larger strengths, in perfect agreement with Ref.~\cite{carb2010}.    

\begin{figure}[h!]
\begin{center}
\includegraphics[width=0.45\linewidth,angle=0,clip=true]{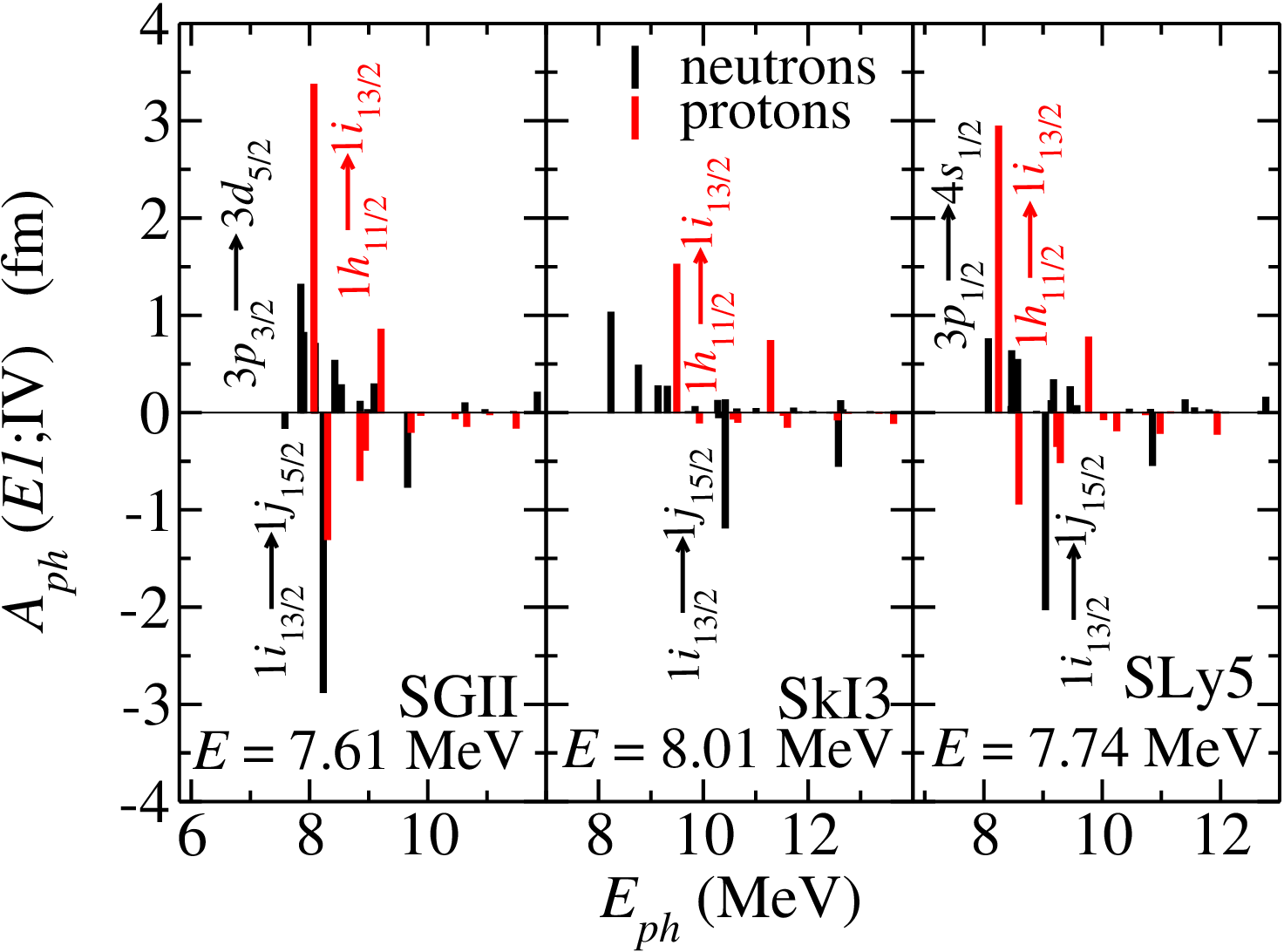}
\includegraphics[width=0.45\linewidth,angle=0,clip=true]{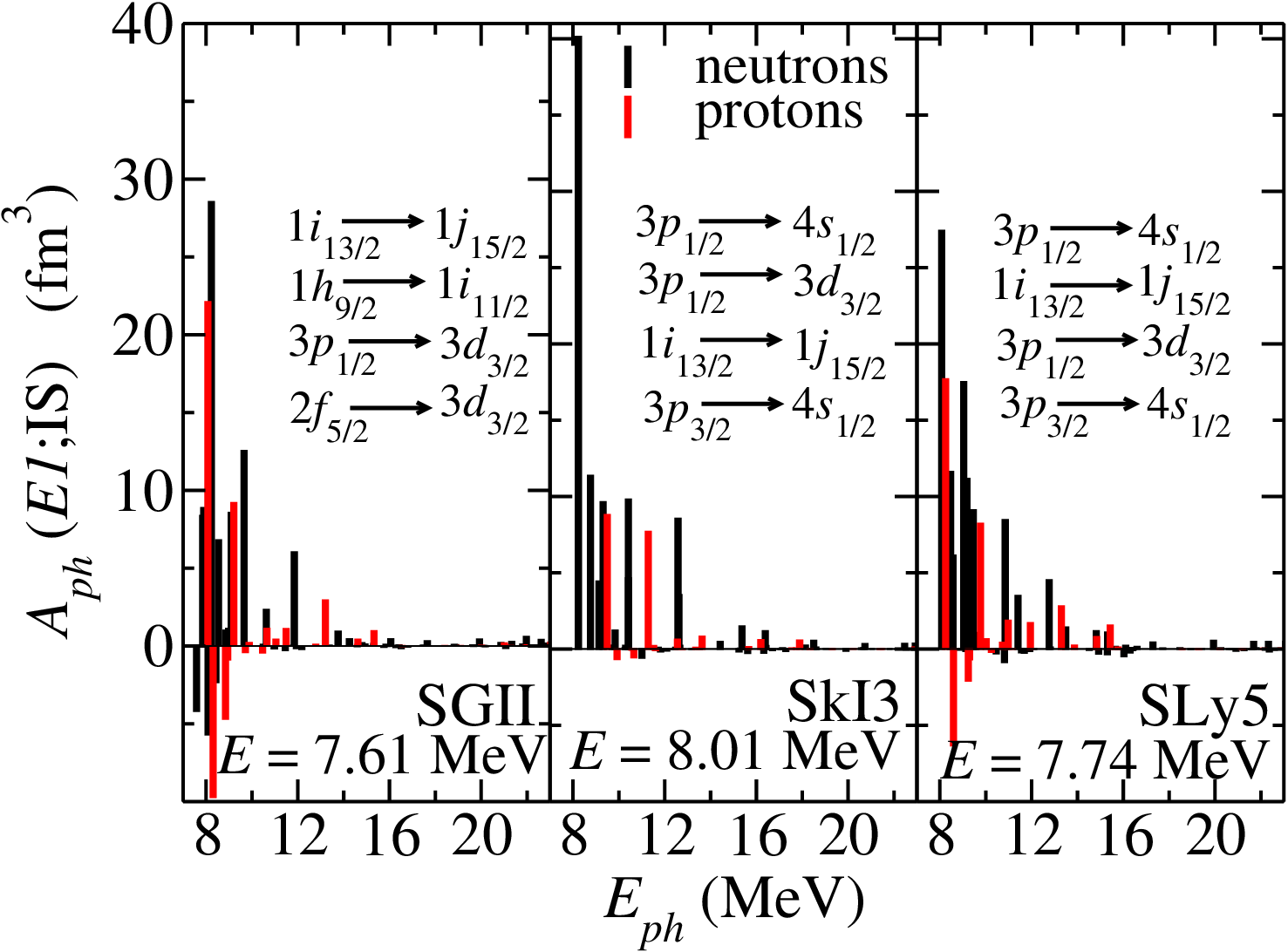}
\end{center}
\caption{\label{fig2} Neutron (black) and proton (red) $ph$ contributions to the isovector (left) 
and isoscalar (right) reduced amplitude [Eq.~(\ref{rts})] corresponding to the {\it RPA-pygmy} state 
of ${}^{208}$Pb (RPA excitation energy is indicated) predicted by the different models as a function of 
the $ph$ excitation energy \cite{pozzi2011}. The single particle levels involved in the most important $ph$ 
transitions are indicated.}
\end{figure}

In Fig.~\ref{fig2}, we show the $ph$ contributions to the IV (left) and IS (right) reduced amplitudes 
[see Eq.~(\ref{rts})] corresponding to the different {\it RPA-pygmy} states of ${}^{208}$Pb predicted by 
the different models as a function of the $ph$ excitation energy. Two results ---common also to other 
mass regions--- arise from the analysis of this quantity. First, while the isoscalar reduced amplitude 
($A_{ph}(E1,IS)$) is formed by several neutron $ph$ transitions adding coherently (same sign) and only 
few states are contributing destructively, the isovector reduced amplitude ($A_{ph}(E1,IV)$) is formed 
by few neutron and proton $ph$ transitions contributing with different signs, i.e., non-coherently.     
Second, the most relevant $ph$ transitions in the isoscalar dipole response of ${}^{208}$Pb correspond 
to excitations of the outermost neutrons that, as a consequence, dominate the dynamics in the low-energy 
region (see \cite{pozzi2011}). For the case of the isovector dipole response the situation is model 
dependent. However, the most relevant neutron $ph$ contributions are larger in number than the proton ones 
and basically due to the outermost neutrons. Therefore, albeit some differences will be present, one may 
expect a relation between the neutron excess and the IS and IV dipole responses in neutron rich nuclei.

From Fig.~\ref{fig1}, we have seen that the low-energy IS and IV dipole responses of ${}^{208}$Pb display, 
to different degree, a sizeable low-energy peak in the corresponding strength functions (see section 
III.B of Ref.~\cite{pozzi2011}). This is actually one of the main characteristics one asks to a collective state. 
The second one is that collective phenomena (or resonances) should display coherence between
the contributions of several $ph$ transitions to the reduced amplitude. Therefore, by looking at Fig.~\ref{fig2}, 
we can state that while all models support a clear collective character of the low-energy peak in the IS 
dipole response in ${}^{208}$Pb (also true in other mass regions), the collectivity of its IV counterpart is 
model dependent. 

\subsection{Parity violating electron scattering on ${}^{208}$Pb}

The Lead Radius Experiment (PREx) at the Jefferson Laboratory has recently
reported first results for PVES on ${}^{208}$Pb \cite{prexIa}. In this first run
statistics were not sufficient in order to achieve the desired accuracy
\cite{prexIa}. A second run of PREx has been approved and it is intended to be
performed in the future \cite{prexIa,prexII}.

\begin{figure}[h!]
\begin{center}
\includegraphics[width=0.45\linewidth,angle=0,clip=true]{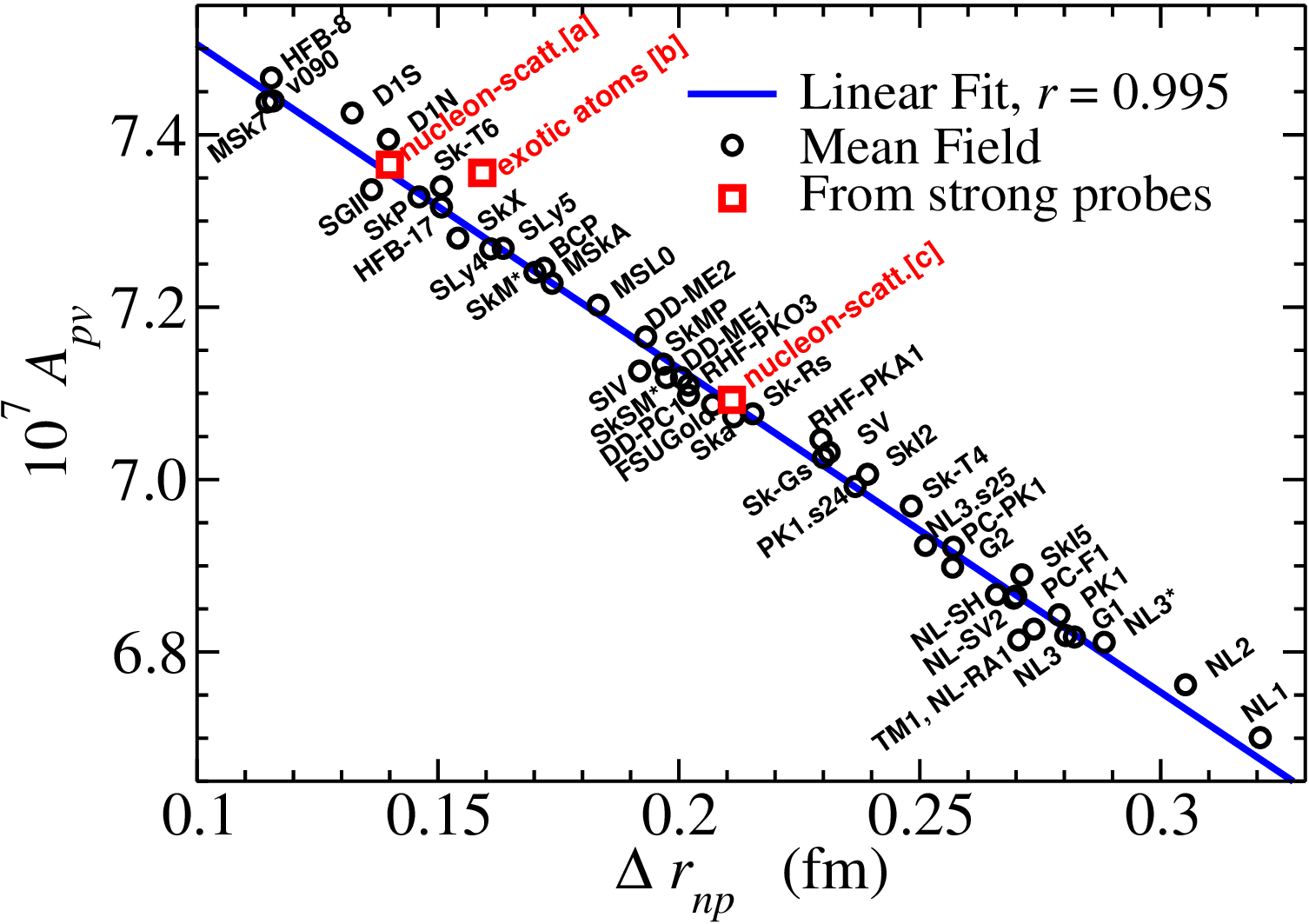}
\includegraphics[width=0.45\linewidth,angle=0,clip=true]{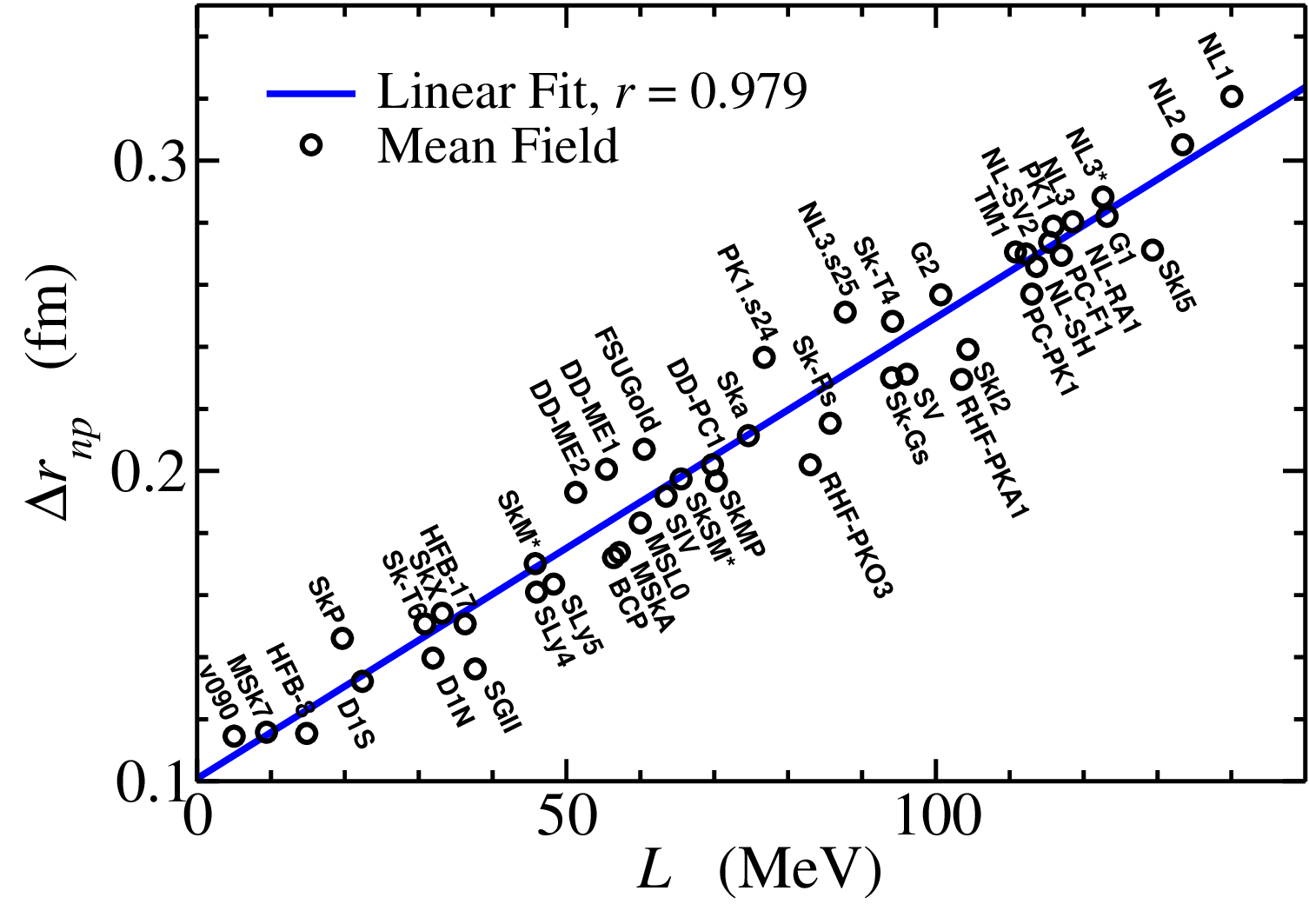}
\end{center}
\caption{\label{fig3} Left panel: parity violating asymmetry (DWBA) 
 in $^{208}$Pb for 1.06~{\rm GeV} electrons at $5^\circ$ scattering angle as 
a function of its neutron skin thickness \cite{roca2011}. 
MF results (black circles) ---references to all these interactions can 
be found in \cite{roca2011}--- and calculations form the neutron densities 
deduced from experiment [a] \cite{hoff1980}, [b]\cite{klos2007} and 
[c] \cite{zeni2010} (red squares). Right panel: Neutron skin thickness 
of $^{208}$Pb as a function of $L$ as predicted by the same MF 
interactions shown in the left panel \cite{roca2011}.}
\end{figure}

As stated in the introduction, the relevance of this measurement has
motivated our study of the parity violating asymmetry at the PREx
kinematics \cite{prexIb,roca2011}. We have applied the formalism mentioned 
in Section \ref{formalism-pves}, which is presented in more detail in 
Refs.~\cite{prexIb,roca2011,horo1998,roca2008b}. In Fig.~\ref{fig3} we
display the linear correlation between the parity violating asymmetry and
the neutron skin thickness of ${}^{208}$Pb 
($\Delta r_{np} = \langle r_n^2\rangle^{1/2} - \langle r_p^2\rangle^{1/2}$) 
as predicted by more than forty mean-field models of very different nature: 
from non-relativistic such as Skyrme or Gogny models to relativistic such as  
non-linear Walecka or density dependent meson-exchange and point-couling models. 
All of them accurately reproduce the charge radius of ${}^{208}$Pb (for further 
details see Ref. \cite{roca2011}). This high linear
correlation allows one to accurately extract the value of the neutron skin
thickness of ${}^{208}$Pb without assuming a particular shape for the nucleon
spatial distributions. In the right panel of Fig.~\ref{fig3} we show the well-known
correlation between the neutron skin thickness in ${}^{208}$Pb and the slope of
the symmetry energy predicted by the considered mean-field models.
Therefore, PVES can supply new constraints on the value of the $L$ parameter. 

\section{Conclusions}
\label{conclusions}

We have exemplified some of the common features found in the dipole response of different 
nuclei \cite{pozzi2011} by showing our results for the case of ${}^{208}$Pb. In particular, 
the low-energy peak in both the isoscalar and isovector dipole responses 
is shifted to larger excitation energies and display larger values of the strength function as 
the value of the $L$ parameter increases. We have 
also seen that the {\it RPA-pygmy} state can be more efficiently excited by an isoscalar 
probe than by an isovector probe. This indicates the dominant isoscalar character of such 
a state. We have demonstrated that the low-energy peak in the isoscalar dipole response 
is a collective mode ---it is formed by several $ph$ contributions adding coherently and 
giving a contribution to the reduced transition strength comparable to that of the isoscalar 
giant dipole resonance. This is found to be opposite to what happens to its isovector 
counterpart where its collectivity depends on the interaction. 
Finally, the dynamics in the low-energy region in the isoscalar dipole response of neutron-rich
nuclei is clearly dominated by the outermost neutrons ---those that form a neutron skin.

Our analysis of PVES applied to the conditions of the PREx experiment
predicts a high-quality correlation between the parity violating asymmetry
and the neutron skin thickness of ${}^{208}$Pb \cite{roca2011}. The results suggest that one
will be able to extract significant constraints on the slope of the
nuclear symmetry energy at saturation if the statistics of PREx are improved. 

In conclusion, the recent experimental and theoretical studies of the
ground-state and excitation properties of neutron-rich nuclei aim to
complement each other and are paving the way for a better knowledge of the
isovector channel of the nuclear effective interaction.

\ack
Work partially supported by grants CSD2007-00042, FIS2008-01661 and 2009SGR-1289 (Spain) 
and DEC-2011/01/B/ST2/03667 from NCN (Poland).

\section*{References}
\bibliography{iopart-num}

\end{document}